\documentstyle[12pt,aaspp4]{article}

\slugcomment{Accepted for publication in the PASP, March 1998} 

\begin{document} 

\title{An Extendable Galaxy Number Count Model} 

\author{Jonathan P. Gardner\footnote[1]{NOAO Research Associate} \\ Laboratory for Astronomy and 
Solar Physics, Code 681, Goddard Space Flight Center, Greenbelt MD 20771 \\Electronic mail: 
gardner@harmony.gsfc.nasa.gov} 

\begin{abstract} 

I review galaxy number count models and present {\em ncmod}, an extendable and general purpose 
model for comparing and interpreting the results of field galaxy survey data. I develop techniques and 
software for converting the results of a survey done in one filter into another filter, for direct comparison 
with other surveys. Comparison of the data from surveys which differ greatly in wavelength coverage or 
sensitivity is of necessity model-dependent, but comparison between similar surveys can be done in a 
relatively model-independent way. I extrapolate existing number counts into the ultraviolet and thermal 
infrared. The model is used to predict the results of future space missions, including STIS and NICMOS 
on HST, ISO, SIRTF and NGST.

\end{abstract} 

\section{Introduction} 

Counting the surface density of galaxies on the sky as a function of apparent magnitude is one of the 
classic cosmological tests. Both the power and the difficulties of extragalactic galaxy surveys were 
recognized by Hubble in the 1930s, when he attempted to constrain the cosmological geometry through 
galaxy counts, but was limited by the difficulty of doing accurate galaxy photometry with the poor 
quality data he had available (Hubble \markcite{hubble34}1934; Hubble \markcite{hubble36}1936), and the inaccuracies of his 
K--corrections. In the 1960s, in a classic review of observational tests of world models possible with the 
Mount Palomar 200--inch telescope, Sandage \markcite{sandage61ne}(1961a) included galaxy counts as one of four 
fundamental tests of observational cosmology. He noted, however, that the number-redshift relation 
$n(z)$ is more sensitive to the value of $q_0$ than the number-magnitude relation $n(m)$. He noted 
further that at the limit of the 200-inch telescope, which he took to be $m_R=22$, the difference in the 
number counts measured in an open universe with $q_0=0$ and a closed universe with $q_0=0.5$ is just 
$\Delta m_R = 0.28$, which was comparable to the photometric and systematic errors of galaxy 
photometry of the time. He claimed that it would be impossible to find the correct world model from 
galaxy counts, a statement that is arguably true even today when we have counts $7 mag$ deeper than the 
limit he considered. To the reasons he listed, systematic and statistical photometric errors, aperture and 
surface brightness corrections, and uncertainties in the idealized models, must now be added the 
unknown details of galaxy formation and evolution. However, some of the more extreme cosmological 
models which he considered, including those with $q_0>1$ or the steady-state model with $q_0 = -1$, 
can be addressed with existing number count data. 

Although also discussed by Sandage \markcite{sandage61ev}(1961b), the effects of luminosity evolution of galaxies on the 
$n(m)$ relation, were not studied in detail until the 1970s. In a series of papers culminating in a review 
of her galaxy count model (Tinsley \markcite{tinsley77}1977; and references therein) developed the technique of 
population synthesis, in which she took an initial mass function (IMF) of stars, and observed broadband 
stellar colors to predict the color and luminosity evolution of stellar populations. To construct her model, 
she made several assumptions which would dominate the study of galaxy counts for the next two 
decades. She assumed that all galaxies formed at the same time, and that the time-scale of star-formation 
was the factor which determined the relation between morphological Hubble type and SED. In this 
picture, elliptical galaxies formed their stars in a short initial burst, while spiral galaxies had longer 
timescales of star-formation. She assumed that there was conservation of galaxy number, and merging 
was not an important factor. 

The population synthesis approach dominated studies of galaxy evolution in the 1980s, when workers 
used stellar spectra instead of broadband colors to construct galaxy templates. In a series of papers 
Bruzual \markcite{bruzual83}(1983a) extended the technique to the ultraviolet (UV) using new observations of stars with the 
IUE telescope. At about the same time, Arimoto \& Yoshii \markcite{arimoto86}(1986), 1987 and Guiderdoni \& Rocca-Volmerange \markcite{guiderdoni87}(1987) included the 
effects of chemical evolution of galaxies, tracing the change in metallicity as the population of stars 
within the galaxy evolves. These results were included in galaxy count models by Yoshii \& Takahara \markcite{yt88}(1988) and 
Guiderdoni \& Rocca-Volmerange \markcite{guiderdoni90}(1990). 

\subsection{Galaxy Count Models} 

The galaxy count model of Yoshii \& Takahara \markcite{yt88}(1988) serves as the basis for this paper, because many of the ingredients 
for the model have now become standard. They began with a measured local luminosity function. They 
assumed that the characteristic magnitude ($M^*$ in the Schechter \markcite{schechter}1976 parameterization) and faint end 
slope ($\alpha$) of the luminosity function was independent of Hubble type in the $b_J$ filter, and set 
the distribution of types (and thus $\phi^*$ as a function of type) to match those observed in a local 
survey. They then converted that luminosity function to other filters using type-dependent local galaxy 
colors from their population synthesis model. This has the effect that while $\alpha$ is the same for each 
type of galaxy in each filter, $M^*$ is a function of type in filters other than $b_J$, and the summed 
luminosity function in other filters has an overall $\alpha$ that is different from that of the individual 
types. Examples of this effect are given below. Other galaxy count modellers (Guiderdoni \& Rocca-Volmerange \markcite{guiderdoni90}1990) used 
type-dependent luminosity functions (e.g. King \& Ellis \markcite{kingellis}1985), even though these were of necessity less 
accurately determined. 

The population synthesis models give the SED of each type of galaxy as a function of time since the 
galaxy formed. (Following Tinsley \markcite{tinsley77}1977, the morphological galaxy types are each equated with a 
star--formation history). The chosen cosmology give the relation between redshift, distance modulus and 
K--corrections, and the time-redshift relation is applied to the luminosity evolution model so as to include 
evolutionary corrections. The model is then integrated over redshift to give the $n(m)$ relation, or 
integrated over other parameters to give $n(z)$, and color distributions. 

Recently a new type of model has appeared. The semi-analytic approach, motivated by cosmological 
theory, was proposed by White \& Frenk \markcite{white91}(1991) and Cole \markcite{cole91}(1991), and developed further by Lacey et al.\ \markcite{lacey93}(1993), 
Kauffman, White \& Guiderdoni \markcite{kauffmann93}(1993), Kauffman, Guiderdoni \& White \markcite{kauffmann94}(1994) and Cole et al.\ \markcite{cole}(1994). This type of theory begins with the physics of 
the big bang, cold dark matter, and hierarchical structure formation, and traces the formation of galaxies 
{\em ab initio} using a semi-analytic approach constrained by numerical simulations. It then predict 
observables, such as the number counts, luminosity functions, colors, redshift distributions, and the 
Tully--Fisher relation. Baugh, Cole \& Frenk \markcite{baughmorph}(1996) have extended this model to study the distribution of 
morphological Hubble types, as measured by the bulge-to-disk ratio. The fundamental difference between 
this type of model and the traditional number count models is the direction in time. Traditional models 
take the local population and attempt to extend it backwards in time, and determine what it evolved from. 
The semi--analytic approach takes the initial conditions set by the big bang and attempts to trace their 
evolution into the local population. While the semi--analytic theory is the most physically motivated, and 
may eventually provide the greatest understanding of galaxy formation and evolution, there is still a place 
for the traditional models. Traditional models are still the best way of comparing different data sets, and 
this is the approach I take in this paper. 

\subsection{Galaxy Count Data} 

A considerable amount of observational galaxy count work was done in the 1980s using photographic 
plates, both on Schmidt telescopes to get bright counts, and on $4m$ class telescopes to get faint counts. 
The modern era of galaxy count data, using linear detectors (CCDs) to count galaxies to very faint levels 
began with the work of Tyson \markcite{tyson88}(1988) and Lilly, Cowie \& Gardner \markcite{lcg}(1991). Those workers established the counts in several 
optical filters to $B\simeq28$, while a series of papers by Metcalfe et al.\ \markcite{metcalfe91}(1991) established CCD counts at 
intermediate magnitudes. These data, when compared to the modeling work described above, led to what 
has become known as the faint blue galaxy problem. The data showed an excess in the number of 
galaxies over the no-evolution model predictions. While the initial interpretation was that the surveys 
were detecting the ultraviolet light from star-forming galaxies in an open cosmology, further 
investigation, including redshift surveys (Broadhurst, Ellis \& Shanks \markcite{bes}1988; Colless et al.\ \markcite{colless}1990; Cowie, Songaila \& Hu \markcite{csh91}1991) showed that this excess 
was not at high redshift. The excess was not seen in the near-infrared $K-$band number counts 
(Gardner, Cowie \& Wainscoat \markcite{gcw}1993), and was made up of blue galaxies. 

The spatial resolution of images taken by the Hubble Space Telescope (HST) has allowed the 
morphological study of faint galaxies. The HST Medium Deep Survey key project (Griffiths et al.\ \markcite{griffiths}1994; 
hereafter MDS) has proven to be a great success in characterizing the optical morphology of galaxies at 
intermediate magnitudes. In conjunction with the fainter images of the Hubble Deep Field 
(Williams et al.\ \markcite{williams}1996; hereafter HDF), the MDS has revealed an excess of galaxies with unusual or disturbed 
morphologies, over what is predicted by the number count models (Glazebrook et al.\ \markcite{kgbmds}1995; Driver, Windhorst \& Griffiths \markcite{drivermds}1995; 
Abraham et al.\ \markcite{abraham}1996). These irregular galaxies likely make up the population of excess faint blue galaxies.

Bright $b_J-$band galaxy counts, measured on scanned photographic plates covering 4300 square 
degrees, show a steeper slope than is predicted by the models (Maddox et al.\ \markcite{maddox}1991). This steep slope was 
initially believed to be due to rapid evolution at low redshift, but recently the data has been called into 
question. Systematics in the photometry, either due to the rapid plate scanning process, 
(Metcalfe, Fong \& Shanks \markcite{metcalfeapm}1995), or to the distribution of central surface brightness in the galaxies, 
(McGaugh \markcite{mcgaugh}1994), could steepen the slope of the number counts artificially. This has caused considerable 
debate about the magnitude of the faint blue galaxy problem. When the models are normalized at the 
bright end of the local steep slope, $B\simeq 15$, the excess in the counts is a factor of $2$ to $4$ over 
the model predictions at $B\simeq22$. The results of redshift surveys of galaxies selected at this 
magnitude level, however, show little departure from no-evolution predictions of the shape of the 
number-redshift relation (Colless et al.\ \markcite{colless}1990), although their absolute numbers, or normalization showed the 
excess seen in the number counts. If the models are normalized at the faint end of the local steep slope, 
$B\simeq18$, and passive evolution is included in the model predictions, then the faint blue galaxy 
excess does not appear until $B>25$ for the closed universe case, which is fainter than any of the 
spectroscopic redshift surveys. Recent bright number counts, obtained with CCD imaging of 10 square 
degrees, do not confirm the steep slope, and are consistent with the higher normalization 
(Gardner et al.\ \markcite{gardnerklfnc}1996). Larger CCD surveys are needed to confirm this result.

There have been several attempts to modify the traditional number count models to explain the faint blue 
galaxy problem by introducing one or more additional free parameters into the model. Rocca-Volmerange \& Guiderdoni \markcite{rocca91}(1990) and 
Broadhurst, Ellis \& Glazebrook \markcite{beg}(1992) relaxed the constraint of conservation of galaxy number, and introduced large amounts of 
merging. A disappearing dwarf galaxy model (Dekel \& Silk \markcite{dekel}1986; Cowie et al.\ \markcite{csh91}1991; Babul \& Rees \markcite{babul}1992) has gained 
support from deep redshift surveys (Lilly et al.\ \markcite{cfrs}1995; Ellis et al.\ \markcite{autofib}1996) which see a steepening of the faint end 
slope of the luminosity function to higher redshift, particularly for the bluer galaxies. High normalization 
models have much less of a faint blue galaxy excess to explain, and succeed in fitting most of the 
observations with modifications to the luminosity evolution models. One recent attempt used a steep 
slope for the luminosity function of late-type spiral galaxies, which has an effect on the counts similar to 
the disappearing dwarf models (Metcalfe et al.\ \markcite{metcalfenat}1996). Cowie et al.\ \markcite{cowie96}(1996) recently proposed a downsizing 
model, in which less luminous galaxies evolve rapidly at low to intermediate redshift, while bright 
galaxies evolve rapidly at redshift $z>1$. 

These models have usually attempted to fit all of the available field galaxy survey data, and do so by 
introducing additional free parameters into the models. Fitting the data, however, is often only done to 
within a factor of $\sim 2$. In this paper I take a more modest approach. Rather than trying to construct a 
single model which fits all of the data, I construct an extendable or general purpose model. The model 
can be tuned to fit one observation, and then used to predict an observation in another filter. This is 
essentially a translation program, translating observations between filters. Number counts and luminosity 
functions measured in different surveys can be directly compared. When the filters are similar, such as 
the ground-based $U$ band filter and the WFPC2 $U_{300}$ filter, this can be done in a relatively 
model independent way, as a model which fits the $U-$band number counts is likely to also fit the 
$U_{300}$ counts. When the filters have very different central wavelengths, such as $B$ and $K$, or 
when the observations are at very different magnitudes, then the comparison will depend more strongly 
on the model parameters chosen. I also will use this program to predict the results of future ground and 
space mission surveys, by extrapolating the most relevant existing data into the filters and depths that will 
be achieved. 

In section 2 I will outline the equations and parameters that make up the model. In section 3 I will use the 
model to compare existing survey data, and to predict the results of future surveys. Section 4 is a 
summary.

\section{The Number Count Model} 

The components which go into a number count model include the spectral energy distributions (SEDs) of 
the galaxies, as a function of galaxy type, either local SEDs to produce a no-evolution model, or SEDs 
that are a function of time to produce an evolution model. Next, a description of the local population of 
galaxies, including the distribution of galaxy types, and the luminosity functions for each type. Finally, 
there is the cosmological model, or the choice of $q_0$, $\Lambda_0$ and $H_0$. The model is then 
integrated. 

\subsection{Spectral Energy Distributions} 

The number count model begins with the spectral energy distribution (SED) of galaxies. The main 
method I use to make this model extendable is in the number and distribution of spectral types of 
galaxies. While previous models have used 3 to 5 spectral types, and equated them to morphological 
Hubble types, (e.g. E/S0, Sa, Sb, Sc, Irr, Yoshii \& Takahara \markcite{yt88}1988), in this model I allow for any number of types, and 
include the ability to interpolate between the types to match an observed color distribution. The 
distribution can be determined by one set of observations, and then the model can be used to predict 
another set of observations. Both theoretical and observed SEDs are available in the literature, the 
theoretical SEDs coming from population synthesis models described above. Bruzual \& Charlot (1996, 
in preparation) have made their galaxy isochrone synthesis spectral evolution library (hereafter 
GISSEL96) model available at {ftp://gemini.tuc.noao.edu/pub/charlot/bc96}, and in Leitherer et al.\ \markcite{leitherer}(1996). This 
model is ideal for the purposes of this paper because it takes the star-formation history of the galaxy and 
returns the SED as a function of time. Any star--formation history is possible, and so it is possible to 
construct a large number of galaxy spectral types. They have included the effects of differing 
metallicities, and find that galaxies have an age--metallicity degeneracy in their colors. 

Kinney et al.\ \markcite{kinney96}(1996) have taken an observational approach to determining the SEDs of galaxies. Using IUE 
pointed and archival data, and ground-based optical spectra on the same galaxies through a similar 
aperture, they have obtained spectra from $1200${\AA} to $10000${\AA} for 30 quiescent and 36 star-
bursting galaxies, with a range of morphological type. By averaging together the spectra of each type, 
they present a library of spectra for 5 quiescent types, the bulge component, and 6 starbursting types. The 
starbursting types are differentiated by their intrinsic color excess $E(B-V)$, caused by dust obscuration, 
which is the dominant parameter that determines the appearance of the starburst SED (Calzetti, Kinney \& Storchi-Bergman \markcite{calzetti}1994). 
The SEDs have been made available at {ftp://ftp.stsci.edu/catalogs/spectra}. 

Most of the Kinney et al.\ \markcite{kinney96}(1996) SEDs extend to $1\mu m$ wavelength, but to match the extensive 
ground-based observations done in the $K$ band at $2.2\mu m$ (e.g. Gardner et al.\ \markcite{gcw}1993), to match the work 
done in the far--IR with IRAS (e.g. Hacking \& Soifer \markcite{hacking91}1991), and to predict the results of planned space 
missions in the near--, mid--, and far--IR, it is necessary to extend the SEDs to other wavelengths. 
Schmidt et al.\ \markcite{schmidt}(1997) have done this, by using previously published data available through the NASA 
Extragalactic Database (NED) to compile the SEDs of the Kinney et al.\ \markcite{kinney96}(1996) galaxy sample over a large 
range in wavelength. While they attempted to correct for the very different apertures used in different 
observations, the processes affecting the SED of galaxies vary widely at different wavelengths, and make 
it difficult to use their SEDs for the determination of K--corrections. Devereux \& Hameed \markcite{devereux}(1997) have shown that 
there is not a simple relation between galaxy type and far--IR SED, and each type of galaxy shows a 
large spread in its ratio of $L_{FIR}/L_{opt}$. My model allows the use of any number of SEDs, and 
allows interpolation between different SEDs, so it is possible to use a range of far--IR fluxes for each 
galaxy template. For the remainder of this paper, I will use SEDs from the GISSEL96 models (which 
include only stellar emission, and thus are not readily applicable in the Far--IR), however, my model can 
be easily extended to include non-stellar emission. This extension will be considered in a future paper.

\subsection{Luminosity Functions} 

There now exist measurements of the luminosity function of field galaxies in the $B$ (Loveday et al.\ \markcite{loveday}1992; 
Marzke, Huchra \& Geller \markcite{marzkelf}1994a), $R$ (Lin et al.\ \markcite{lin}1996), and $K$ (Gardner et al.\ \markcite{gardnerklf}1997) bands, and in the far--IR 
(Saunders et al.\ \markcite{saunders90}1990; Isobe \& Feigelson \markcite{isobe91}1991). In the optical and near--IR, a Schechter \markcite{schechter}(1976) parameterization is 
often used, while in other bands, a Gaussian or power--law/Gaussian combination is used. The 
Schechter \markcite{schechter}(1976) parameterization is 
\begin{equation} 
\phi (L)dL = {\phi}^* (L/L^*)^{\alpha} exp (-L/L^*)d(L/L^*) , 
\end{equation}
\noindent or equivalently  
\begin{equation} 
\label{lumfn} 
\psi (M)dM = \frac{ln(10)} {2.5} {\phi}^* exp \{ -\frac{ln(10)} {2.5}(\alpha + 1)(M-M^*) - exp[ -
\frac{ln(10)} {2.5}(M-M^*)] \} dM , 
\end{equation}  
\noindent where $exp(x)=e^x$, $ln(x)$ is the natural logarithm, $\alpha$ is a constant index, ${\phi}^*$ 
is a constant coefficient which has the dimension of the number density of galaxies, $L^*$ is the 
characteristic luminosity, and $M^*$ is the corresponding characteristic magnitude. Generally, the 
measurement of a luminosity function is also done in a non-parametric manner (Efstathiou, Ellis \& Peterson \markcite{eep}1988) and a discrete 
luminosity function is returned.

There have been several determinations of the local luminosity function of galaxies measured as a 
function of type (Loveday et al.\ \markcite{loveday}1992; Marzke et al.\ \markcite{marzke94}1994b), color (Metcalfe et al.\ \markcite{metcalfe91}1991; Shanks \markcite{shanks90}1990; 
Marzke \& da Costa \markcite{marzke97}1997), or emission line properties (Lin et al.\ \markcite{lin}1996), as well as studies of the evolution of the LF 
with redshift (Lilly et al.\ \markcite{cfrs}1995; Ellis et al.\ \markcite{autofib}1996; Cowie et al.\ \markcite{cowie96}1996). The consensus of these studies is that red, 
early-type, and non-emission-line galaxies tend to have a shallower slope, (i.e. $\alpha > -1.0$), than the 
blue, late-type, and emission-line galaxies ($\alpha < -1.0$). While there is a consensus on the trend, 
however, there is not a consensus on the actual measurement of $\alpha$, which is highly correlated with 
$M^*$ in any case. I have included the capability within my model to use a different LF for each spectral 
type.

\subsection{Cosmological Parameters} 

Number count models are constructed in co-moving coordinates, as that is how luminosity functions are 
formulated. Following Yoshii \& Takahara \markcite{yt88}(1988), the relation between apparent magnitude and absolute magnitude in a 
filter $F$ is: 
\begin{equation} 
\label{mvz} m_{F} = M_{F} + E_{F}(z) + 5 log (D_L/10 pc) , 
\end{equation}  
\noindent where $log$ is the base 10 logarithm,  
\noindent and  
\begin{equation} 
\label{ecorr} E_{F}(z) = -2.5 log \frac {\int_0^{\infty} f_{{\lambda}^{\prime} / (1+z)} (t_G(z)) F 
({\lambda}^{\prime}) d{\lambda}^{\prime} /(1+z)} {\int_0^{\infty} f_{{\lambda}^{\prime} } (t_G(0)) 
F ({\lambda}^{\prime}) d{\lambda}^{\prime} } , 
\end{equation}  
\noindent where $F ({\lambda})$ is the filter throughput. In the no-evolution case, $E_{F}(z)$ is 
evaluated at $t_G(0)$, and is called the K--correction.

In the Friedman--Robertson--Walker model, the luminosity distance $D_L$ is defined as (Carroll, Press \& Turner \markcite{carroll}1992; 
Fukugita et al.\ \markcite{fukugita}1990):  
\begin{equation} 
\label{lumdist} D_L = \frac {c(1+z)} {H_0 |\Omega_k|^{1/2}} sinn \{|\Omega_k|^{1/2} 
\int_0^z [(1+z^{\prime})^2(1+2(q_0+\lambda_0)z^{\prime}) - z^{\prime}(2+z^{\prime})\lambda_0]^{-
1/2}dz^{\prime} 
\} , 
\end{equation}  
\noindent where $H_0$, $q_0$, and $c$ are the Hubble constant, the deceleration parameter, and the 
speed of light, respectively, and $\lambda_0$ is the normalized cosmological constant, $\lambda_0 
\equiv 
\Lambda_0 c^2/3H_0^2$. $\Omega_k$ is the curvature term, $\Omega_k = 1- 2q_0-3\lambda_0$, and 
$sinn$ is defined as $sinh$ for models with $\Omega_k > 0$ (open universe) and $sin$ for models with 
$\Omega_k<0$ (closed universe). In a flat universe, $\Omega_k=0$, and the $sinn$ and $\Omega_k$s 
disappear from the equation, leaving only the integral. The comoving volume $V(z)$ and the 
cosmological time $t(z)$ differentiated with respect to $z$ are given by 
\begin{equation} 
\label{dvdz} 
\frac {dV} {dz} = \frac {4 \pi c D_L^2} {H_0 (1+z)^2 [2(q_0+\lambda_0)(1+z)^3 + \lambda_0 + 
\Omega_k (1+z)^2]^{1/2} } , 
\end{equation}  
\noindent and  
\begin{equation} 
\label{dtdz} 
\frac {dt} {dz} = \frac {-1} {H_0 (1+z) \{[2(q_{0}+\lambda_0)z+1- 
\lambda_0](1 + z)^2 + \lambda_0\}^{1/2} } , 
\end{equation}  
\noindent respectively. The age $t_G(z)$ of a galaxy formed at redshift $z_F$ is obtained by integrating 
$dt/dz$ from $z_F$ to $z$. $H_0$ cancels out in this model in every factor except $t_G(z)$, and thus has 
no effect on no-evolution models. In models with evolution, a change in $H_0$ is degenerate with a 
change in $z_F$, as long as the current age of the galaxies is less than the Hubble time. 

\subsection{Constructing the Model} 

The galaxy number count data are obtained by counting all the images of galaxies in a finite area of the 
sky. If $n(m_{\lambda},z)dm_{\lambda}dz$ is the number of galaxies between $m_{\lambda}$ and 
$m_{\lambda}+dm_{\lambda}$ and between $z$ and $z+dz$, then for ($0 \leq z \leq z_F$), 
\begin{equation} 
\label{nmz} n(m_{\lambda},z)dm_{\lambda}dz = \frac {\omega} {4 \pi} \frac {dV} {dz} \psi 
(m_{\lambda},z) dm_{\lambda} dz , 
\end{equation}  
\noindent where $\omega$ is the angular area in units of steradians over which the galaxies are counted, 
and $\psi (m_{\lambda},z)$ can be obtained by solving equation~\ref{mvz} for $M_{\lambda}$ (for 
each galaxy type) and putting the result into equation~\ref{lumfn}. I will use $1 ~ deg^2$ for our area, 
which corresponds to $\omega = 3.05 \times 10^{-4} sr$.  Integrating 
$n(m_{\lambda},z)dm_{\lambda}dz$ with respect to $z$, gives the differential number count  
\begin{equation} n(m_{\lambda})dm_{\lambda} = \int_0^{z_F} n(m_{\lambda},z)dm_{\lambda}dz , 
\end{equation} 
\noindent which is summed over the distribution of galaxy types. 

\subsection{Model dependence} 

\begin{figure}

\plotone{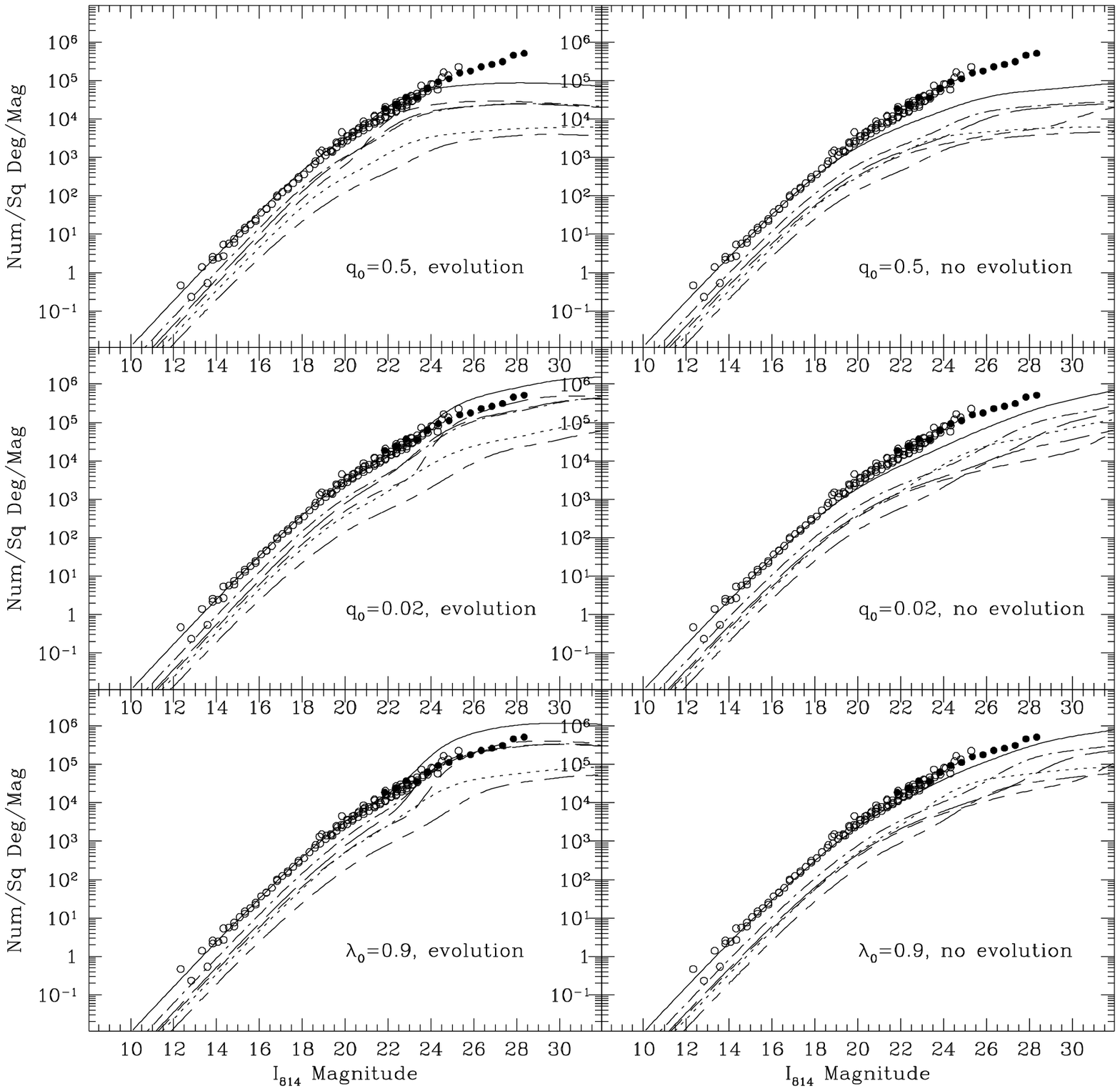}

\caption{The dependence of the fiducial $I_{814}$ model on cosmological geometry and evolution. In 
each panel I plot the total number counts of galaxies (solid line), and the number counts of each of 5 
types: E/S0 (short dashed), Sab (long dashed), Sbc (dash-dot), Scd (short dash-long dash) and Irr (dotted). 
Also plotted are a compilation of the $I-$band data (open points) including the HDF  (filled points). The 
panels on the left include the effects of passive luminosity evolution, the panels on the right are 
no-evolution, or pure K--correction models.}

\label{ibase}

\end{figure}

There are a considerable number of free parameters in this model, although some can be determined from 
local survey observations. In this section I consider the effects on the model of varying the free 
parameters one at a time, while holding the other parameters fixed. The fiducial model I will use for 
comparison is a model of the number counts in the WFPC2 $I_{814}$ filter, for a model with $q_0=0.5$, 
$H_0=50 ~km~sec^{-1}~Mpc^{-1}$, and $\lambda_0=0.0$, including the effects of evolution. I 
consider the effects of varying the galaxy mix below, so in this section I will use a galaxy mix consisting 
of 4 types with a single redshift of formation, and exponential rates of star formation with e-folding times 
of 1 Gyr, 4 Gyr, 7 Gyr, and constant star formation, corresponding roughly to E/S0, Sab, Sbc and Scds. 
Half of the E/S0 types have metallicities of 2.5 times solar, the other half have solar metallicity. The 
Spiral galaxies have solar, 2/5, and 1/5 solar metallicity respectively. The galaxies were formed at 
$z_{form}=15$, and are in a distribution of 32\%, 28\%, 29\%, and 5\%, respectively. I include one 
additional starforming dwarf type, with constant starformation, 1/5 solar metallicity, and an age of 1 Gyr 
at every redshift. This last type does not passively evolve, and represents a population of starforming 
galaxies that is continuously refreshed at all redshifts (Gronwall \& Koo \markcite{gronwall}1995; Pozzetti, Bruzual \& Zamorani \markcite{pozzetti}1996). For this fiducial 
model, I have used the type-independent $K-$band luminosity function as measured by Gardner et al.\ \markcite{gardnerklf}(1997), 
($M^*_K = -24.62$, $\alpha = -0.91$, and $\phi^* = 2.08 \times 10^{-3}$), converted to the 
type-dependent $I_{814}$ band as described above. The models are plotted in Figure~\ref{ibase}, along 
with a compilation of the $I-$band data. The galaxy types are plotted separately to show the contribution 
of each type to the total.

In the past, number count model predictions have often been plotted as a function of cosmology, and with 
and without the effects of evolution. I do this in Figure~\ref{ibase}. There are several aspects of this 
fiducial model to note. First, the no-evolution models underpredict the number counts at the faint end in 
all of the cosmologies plotted. Second, evolution causes the model to shoot up rapidly at $I_{814} 
\approx 20$ for the $q_0=0.5$ case, and at $I_{814} \approx 22$ for the other two cosmologies. This 
effect, although quickly swamped by volume effects in the $q_0=0.5$ model, results in an overprediction 
of the observed counts, and is due to the rest-frame ultraviolet flux of the massive stars involved in the 
initial burst of star-formation of the galaxies. This overprediction was absent from many early models. 
Some were based upon the tabulated Bruzual \markcite{bruzual83m}(1983b) models which only extended to $z=2$, and thus 
did not include the effects of this initial burst of star-formation. Others used a parametric form for 
star-formation (i.e. Tinsley \markcite{tinsley77}1977), which also under-estimated the initial burst of star-formation. 
Wang \markcite{wang}(1991) proposed that extinction of the UV light by dust produced by the star-formation (which is not 
included in the GISSEL96 models) will eliminate this overprediction of the models (Gronwall \& Koo \markcite{gronwall}1995; 
Babul \& Ferguson \markcite{babul96}1996; Roche et al.\ \markcite{roche}1996). I consider this effect below. In addition, Pozzetti et al.\ \markcite{pozzetti}(1996) used a Scalo \markcite{scalo}(1986) 
initial mass function, rather than a Salpeter \markcite{salpeter}(1955) IMF to reduce the production of UV light.

\begin{figure}

\plotone{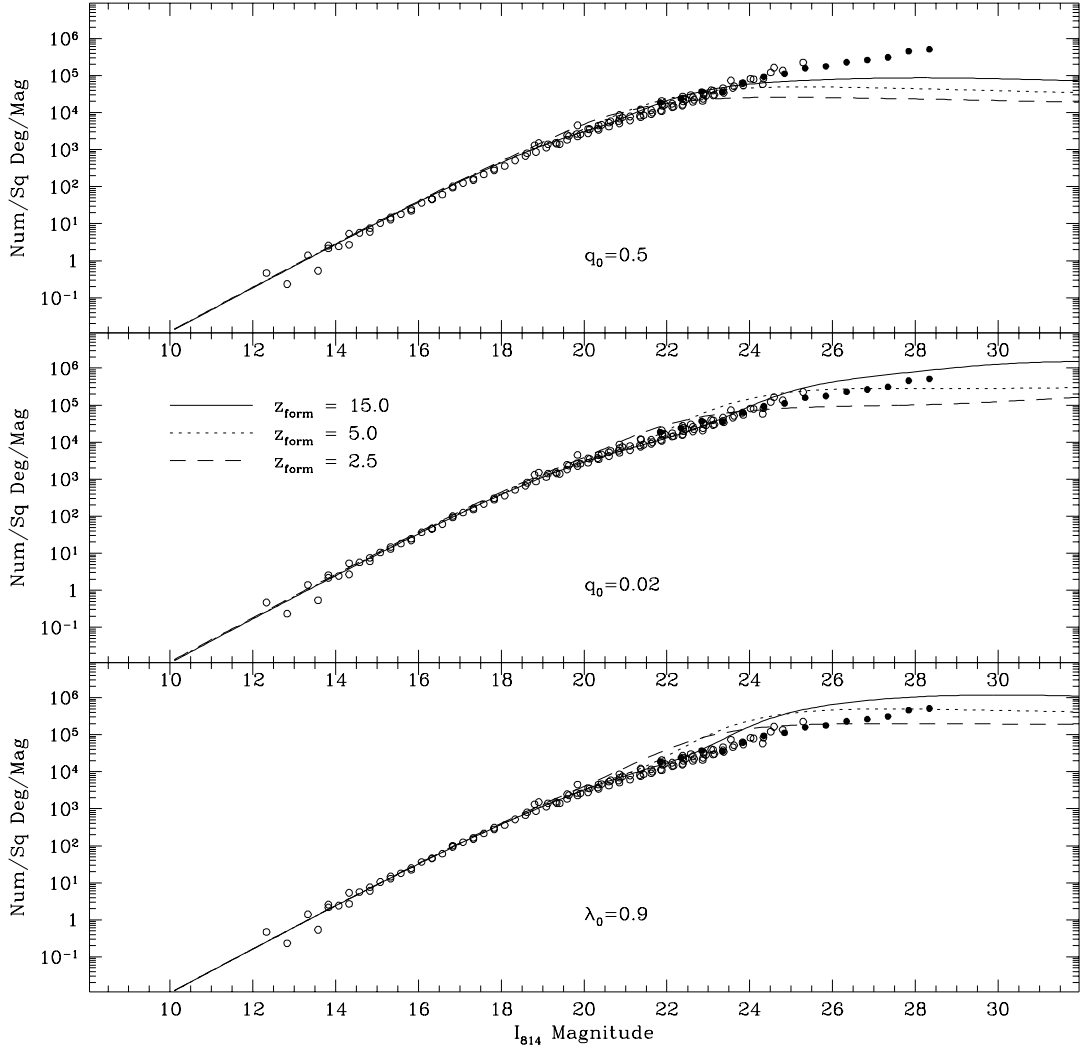}

\caption{The dependence of the models on the redshift of galaxy formation. The data and the 
$z_{form}=15$ models are the same as the evolution models in Figure~\protect\ref{ibase}, and I have 
included models with $z_{form}=5$ and $z_{form}=2.5$. The effect of a lower redshift of galaxy 
formation is to introduce a turnover in the counts. The faint end of the counts is then dominated by the 
slope of the faint end of the luminosity function.}

\label{zform}

\end{figure}

The models plotted in Figure~\ref{ibase} each have $z_{form} = 15$, and I now consider the effects on 
the models of varying the redshift of galaxy formation. In Figure~\ref{zform}, I plot the results of the 
models run with $z_{form}=5$ and $z_{form}=2.5$. The redshift of formation causes a turnover in the 
number counts. At fainter magnitudes, the surveys are no longer reaching to higher redshift, and therefore 
the number counts are no longer going up due to volume effects. The limiting slope of the number counts 
is set by the faint end slope of the luminosity function. This has led some workers to propose that there is 
a component of the local luminosity function with a steep slope, which comes to dominate at the faint 
end (Driver et al.\ \markcite{driver94}1994a). There is evidence that dwarf elliptical galaxies have a steep slope in the luminosity 
function measured in some clusters (Sandage, Binggelli \& Tammann \markcite{sandagevirgo}1985; Driver et al.\ \markcite{driverclus}1994b). In the field there have been 
attempts to detect this population, (Marzke et al.\ \markcite{marzkelf}1994a; Loveday \markcite{loveday97}1997), but it is not yet clear whether it is 
a universal phenomenon, or a feature of overdense regions.

\subsection{Internal Absorption by Dust}

\label{dustsect}

In this section I consider the effects on the models of absorption by dust internal to the galaxies. First 
investigated by Wang \markcite{wang}(1991), it was recognized that the population synthesis models predicted too much 
UV flux in the absence of dust (Gronwall \& Koo \markcite{gronwall}1995; Roche et al.\ \markcite{roche}1996; Babul \& Ferguson \markcite{babul96}1996). Following 
Bruzual, Magris \& Calvet \markcite{bruzualdust}(1988), Wang \markcite{wang}(1991) modeled the dust internal to a galaxy as an absorbing layer, symmetric 
around the midplane of the galaxy, whose thickness is a fraction $\zeta$ of the total thickness of the 
stellar disk. The observed luminosity is 
\begin{equation}
L_{observed} = L[(1-\zeta)(1+e^{-\tau})/2 + \zeta(1-e^{-\tau})/\tau] ,
\end{equation}
\noindent where $\tau$ is the optical depth of the absorbing layer. I take $\zeta = 0.25$, and assume that 
the optical depth $\tau \propto L^{\beta}$, where $\beta = 0.5$. I will assume that the extinction is a 
power law in wavelength, $\propto \lambda^{-n}$, with $n=2$. I set the present day extinction of 
galaxies to be $\tau = 0.2$ for $L^*$ galaxies at $4500${\AA}, and scale all other galaxies according to 
the above equations. Using the assumption that the measured luminosity function has not been corrected 
for the effects of internal absorption by dust, the rest-frame correction is subtracted.

\begin{figure}

\plotone{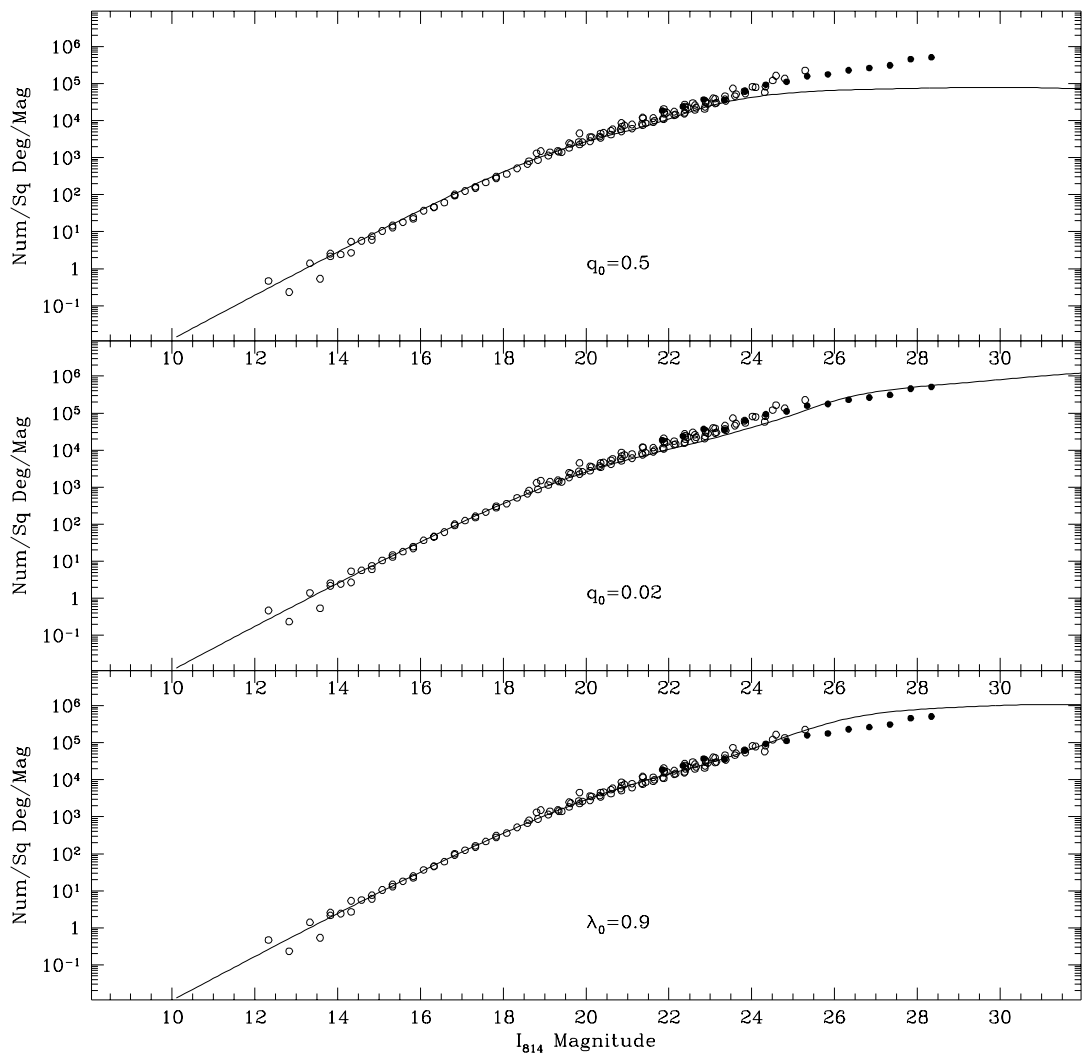}

\caption{The effects on the models of internal absorption by dust. The models plotted are the same as the 
evolution models in Figure~\protect\ref{ibase}, but now the effects of internal absorption by dust are 
included. The dust absorption reduces the strong UV flux that is predicted for forming galaxies by the 
population synthesis models, and eliminates the strong up-turn in the counts seen in the dust-free models 
at intermediate magnitudes.}

\label{dustfig}

\end{figure}

Figure~\ref{dustfig} is a plot of the model prediction for the $I_{814}$ number counts, including the 
effects of dust. In comparison to the models plotted in Figure~\ref{ibase}, the models no longer 
overpredict the counts for the $\lambda_0 = 0$ models, as the UV flux from the first burst of star 
formation is no longer seen at bright magnitudes. Further evidence for the role of dust in galaxy 
formation comes from long-slit spectroscopy and narrow-band imaging searches for line emission from 
starbursts at high redshift (Thompson \& Djorgovski \markcite{thompson}1995), and from models of damped $Ly\alpha$ 
systems(Pei \& Fall \markcite{pei}1995).

\subsection{Merging, or Galaxies in Pieces}

Following suggestions by Koo \markcite{koo90}(1990) and Guiderdoni \& Rocca-Volmerange \markcite{guiderdoni90}(1990), Rocca-Volmerange \& Guiderdoni \markcite{rocca91}(1990) proposed that the 
galaxies in the present-day universe were formed in pieces which subsequently merged together. In their 
model, this number evolution takes the form $\phi^* \propto (1+z)^{\eta}$ in the Schechter 
parameterization, and in order to conserve the luminosity density, $L^* \propto (1+z)^{-{\eta}}$; here, 
$\eta$ is a free parameter. This has the effect of dividing the flux from a single galaxy at high redshift 
between several galaxies. While this is usually considered to be due to the physical merging of galaxies, 
it also would be the result of an observational artifact introduced into the data by over-enthusiastic 
deblending in the object detection routine. Colley et al.\ \markcite{colley}(1996) recently proposed that this observational effect is 
operating in the catalogs of the HDF, and is caused by the fact that galaxies tend to be clumpy and 
irregular in the rest-frame UV. Broadhurst et al.\ \markcite{beg}(1992) proposed a slightly more complicated function of redshift, 
$\phi^* \propto exp(-Q/\beta((1+z)^{-\beta}-1))$, where $Q$ defines the merger rate, and $\beta$ is a 
function of the look-back time. This function avoids the unreasonably high merger rate at high redshift of 
the exponential form, and has the intuitive advantage that $Q$ is approximately the number of pieces at 
$z \approx 1$ that merge to form a present-day galaxy. Glazebrook et al.\ \markcite{kgbnc}(1994) used a simplified version of this 
function, with the merger rate $\propto 1+Qz$. Guiderdoni \& Rocca-Volmerange \markcite{guiderdoni91}(1991) give a detailed analysis of 
self-similar merging scenarios with the assumption of conservation of total comoving luminosity density, 
which leads to:
\begin{equation}
L^*(z)\phi^*(z)\Gamma(\alpha(z)+2) = L^*(0)\phi^*(0)\Gamma(\alpha(0)+2),
\end{equation}
\noindent where $\Gamma(\alpha+2)$ is the Gamma function that results from integrating the Schechter 
luminosity function. They alternatively hold each of the Schechter parameters constant while varying the 
other two. In this paper I will only consider merging models with $\alpha$ held constant, but the model is 
easily modified for the case when $\alpha$ is a function of redshift. Both Broadhurst et al.\ \markcite{beg}(1992) and 
Carlberg \markcite{carlberg92}(1992) include in their models the effects of merging on the luminosity evolution of the 
galaxies, but I will not consider that here.

\begin{figure}

\plotone{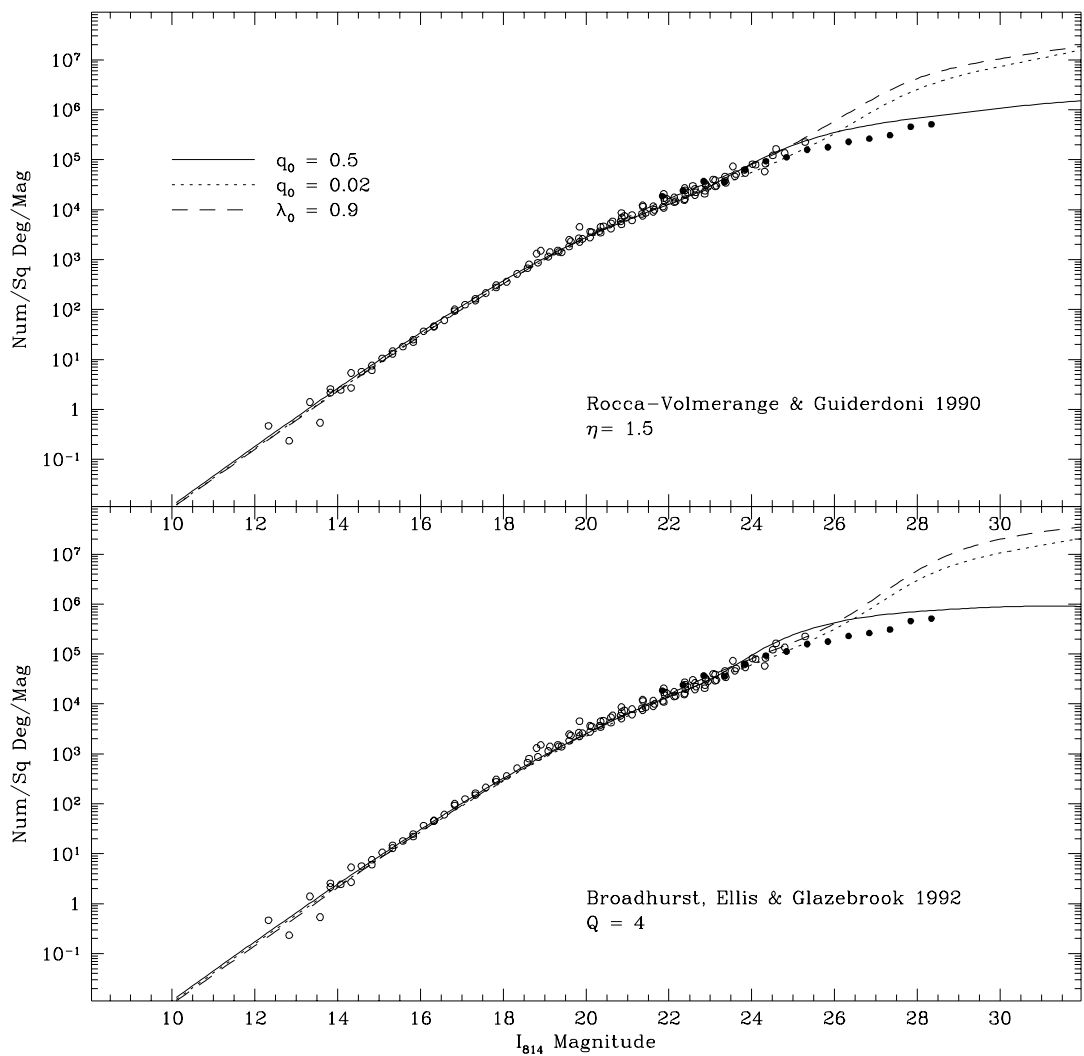}

\caption{The effects of merging on the models. The models include self-similar merging with 
conservation of luminosity density. Merging has the effect of lowering the luminosity of individual 
galaxies while increasing the numbers at fainter magnitudes. The high observed counts at faint 
magnitudes do not rule out the $q_0=0.5$, $\lambda_0=0.0$ model if merging is included. The top panel 
uses the formulation of Rocca-Volmerange \& Guiderdoni \protect\markcite{rocca91}(1990), while the bottom panel uses the formulation of 
Broadhurst et al.\ \protect\markcite{beg}(1992).}

\label{merge}

\end{figure}

Figure~\ref{merge} is a plot of the model predictions for the fiducial model with Rocca-Volmerange \& Guiderdoni \markcite{rocca91}(1990) merging, 
and with Broadhurst et al.\ \markcite{beg}(1992) merging. The two models give very similar results. In general, the decrease in 
luminosity of individual galaxies moves to fainter magnitudes the sharp upturn in the counts due to the 
UV flux from the initial burst of star-formation. The increase in number density causes the counts at 
fainter magnitudes to rise. The $q_0=0.5$, $\lambda_0=0.0$ cosmology is no longer ruled out by the 
high counts at the faint end.

\subsection{Excess Dwarf Models}

\begin{figure}

\plotone{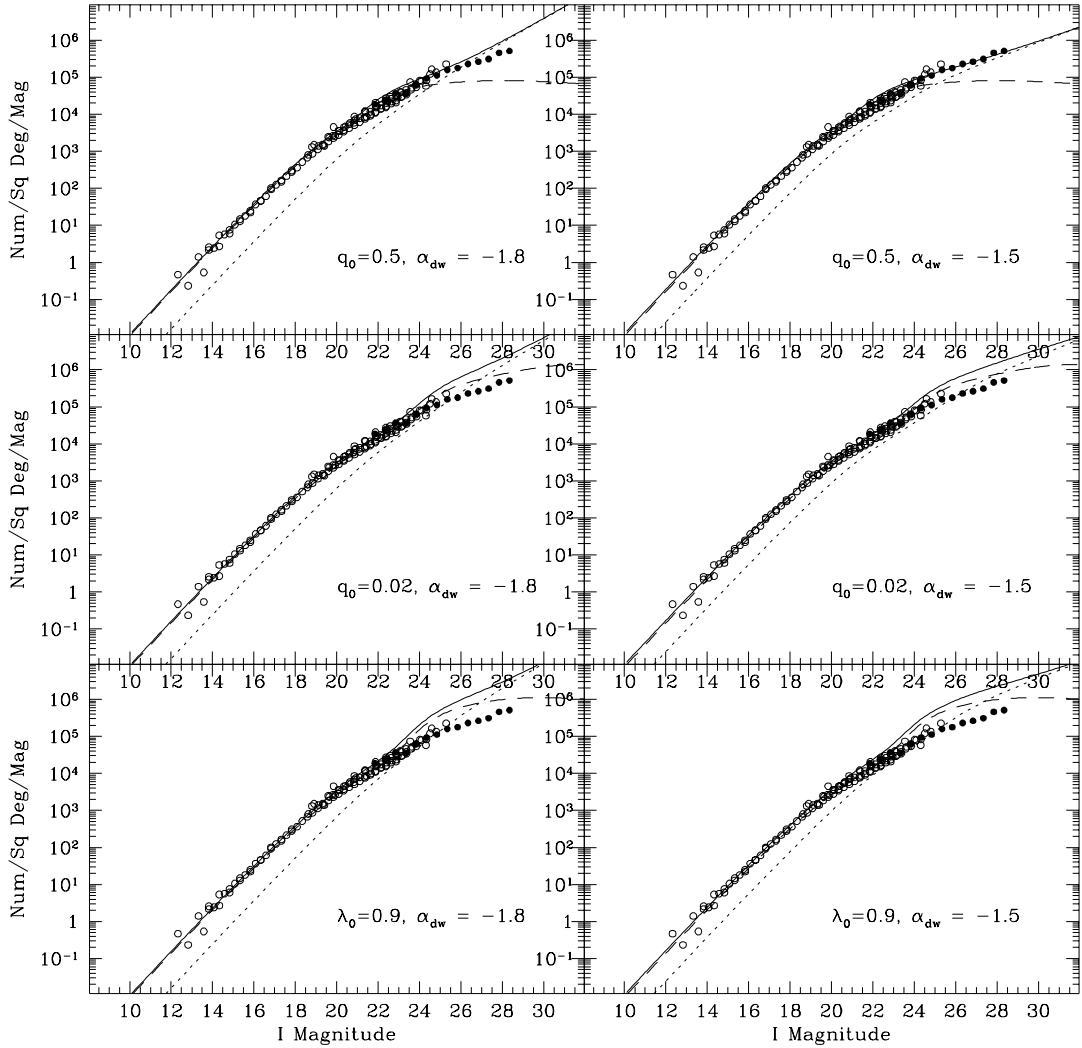}

\caption{The effects of an extra population of dwarf galaxies on the models. In each panel, the solid line 
is the total number of galaxies, the dotted line is the number of dwarf galaxies, and the dashed line is the 
rest of the galaxies. A population of dwarf galaxies with a steep faint end slope has very little effect on 
the bright number counts, but comes to dominate at faint magnitudes. This is because the large galaxies 
with faint apparent magnitudes are at high redshift, where there is little volume and thus low numbers, 
while the dwarf galaxies are at low redshift where there is greater volume.}

\label{dwarffig}

\end{figure}

Number count models with an excess of dwarf galaxies have also been proposed as a way of reconciling 
the high counts with a $q_0=0.5$, $\lambda_0=0.0$ cosmology. They have also been used to explain the 
number-redshift relation (Cowie et al.\ \markcite{csh91}1991), and the observed excess of irregular galaxies seen in the HST 
Medium Deep Survey (Glazebrook et al.\ \markcite{kgbmds}1995; Driver et al.\ \markcite{drivermds}1995), and in the Hubble Deep Field (Abraham et al.\ \markcite{abraham}1996). 
The dwarf galaxies in these models are actively star-forming, a requirement introduced to fit the increase 
in emission line strength seen in the spectroscopic redshift surveys. I include this type of galaxy 
population in my model, and plot the results in Figure~\ref{dwarffig}. I give the constant star-formation 
galaxy type a steep slope $\alpha=-1.8$ in the luminosity function, balanced by a fainter $M^*_K = -
23.12$, and $\phi^*=7.5 \times 10^{-3}$. The dwarf galaxies have a negligible effect at bright 
magnitudes because of the faint $M^*$, but dominate the number counts at faint levels due to the steep 
faint end slope. This model is somewhat {\em ad hoc}, as are many of the modifications to the standard 
models. It allows any arbitrary faint-end slope for the number counts, with the only stipulation being that 
the local dwarf population be too faint to be observed in the local redshift surveys used to determine the 
luminosity function. The faint end slope of the number counts is not as steep as this model predicts, so I 
also plot a model where the dwarf galaxies are given a faint end slope of $\alpha = -1.5$.

\begin{figure}

\plotone{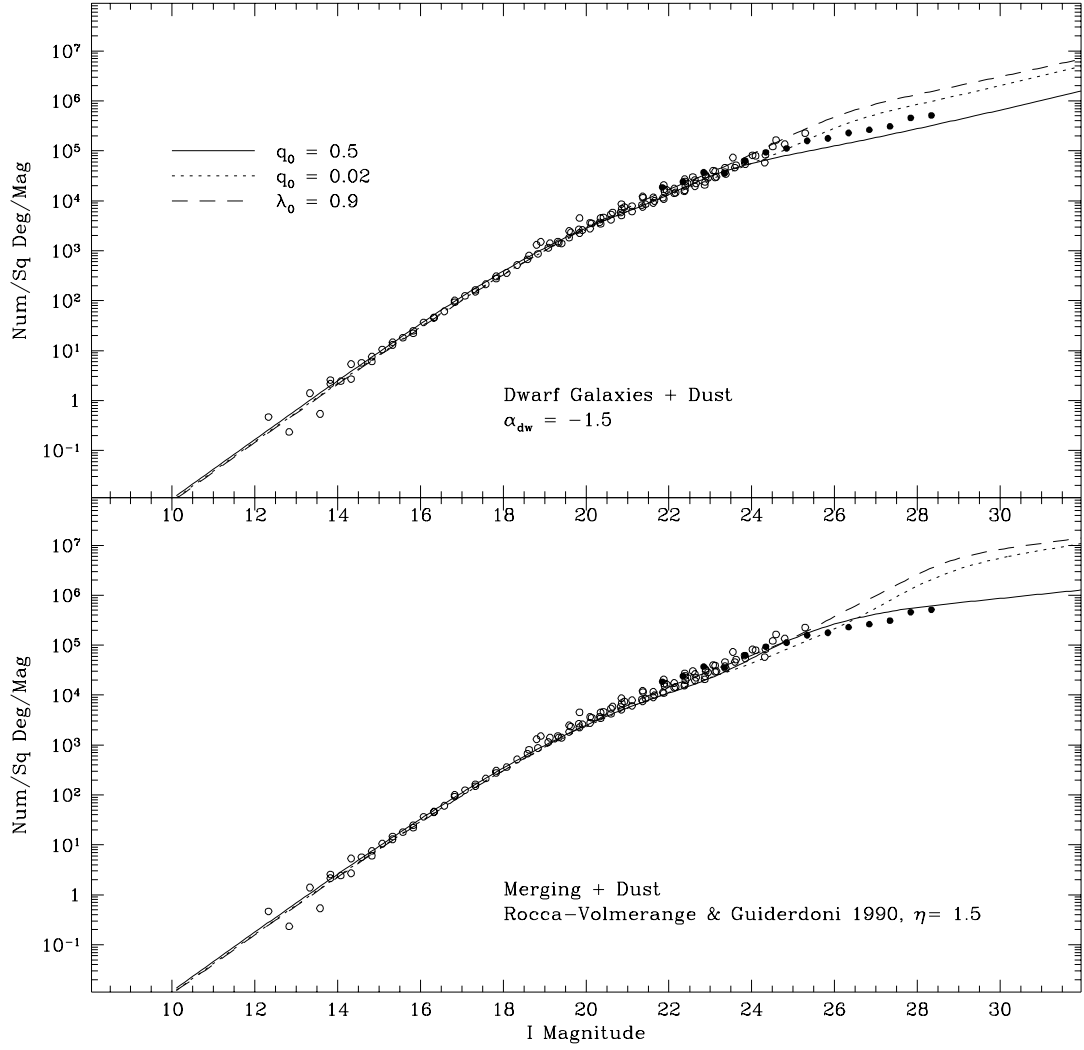}

\caption{The combined effects of dwarf galaxies plus dust, and merging plus dust on the models.}

\label{dwmgdust}

\end{figure}

Figure~\ref{dwmgdust} is a plot of the model predictions for dwarf galaxies plus dust, and for 
Rocca-Volmerange \& Guiderdoni \markcite{rocca91}(1990) merging plus dust. This figure illustrates the interaction between the free parameters of 
the model.

\section{Applications of the Model}

In the last section I reviewed the ingredients of a traditional galaxy count model. By including all of the 
extra free parameters that have been invoked to fit the data, I have made it possible to fit observations of 
galaxy counts in a variety of different ways. I have shown that contrary to some early hopes, it is not 
realistic to expect galaxy counts alone to constrain the cosmological geometry, or even to strongly 
constrain the form of galaxy evolution. While some values of the free parameters are physically more 
realistic than others, nonetheless there is too much degeneracy in the possible models. Surveys, however, 
contain much more data than just the number of galaxies. Most photometric survey fields have been 
observed with more than one filter, giving color information on the galaxies. Galaxy evolution is 
measured by studying the change with redshift of the general luminosity function, and spectroscopic 
redshift surveys provide the greatest leverage for these studies. Models like the one I have presented here 
can be used to directly interpret survey results.

My purpose in this paper is not to construct a single model which fits all the data, however. The power of 
galaxy count models like this one is in comparing different observations with each other. The simplest 
application is a consistency check: e. g. is the local $b_j-$band luminosity function, as measured by 
Loveday et al.\ \markcite{loveday}(1992), consistent with the faint $B-$band number counts, as measured by Tyson \markcite{tyson88}(1988). 
Answering this question in the negative led to the faint blue galaxy problem, and a modification of the 
models. In this section I will extend this use of the model to comparing observations made in different 
filters. I will use the model to convert measurements of the galaxy counts, and of the luminosity functions 
from one filter to another. This will serve two purposes. First, observations made in different filters can 
be directly compared with one another. Second, it will be possible to make predictions for the results of 
future observations using new filters, especially for future space missions that will address wavelength 
regimes for which there is very little existing data. 

\subsection{Conversion between filters} 

Comparing the results of a survey conducted in one filter to a survey conducted in a different filter can be 
troublesome. While previously converting number counts between similar filters has been done with a 
constant offset in magnitude (Metcalfe et al.\ \markcite{metcalfe91}1991), this doesn't work for comparing observations between 
filters that are different enough that the slope will change. It is possible, however, to use a number count 
model to translate or interpolate measured counts from one filter to another. This is of necessity 
model-dependent, but when translating counts between very similar filters, such as from the Johnson 
$U_{J}$ filter to the WFPC2 $U_{300}$ filter, or from the ground-based $K$ filter to the NICMOS 
$H_{1.6}$ filter, the choice of model parameters makes little difference. When translating counts 
between filters with widely separated central wavelengths, or when extrapolating ground-based counts 
into the UV or thermal IR, the resulting prediction depends more strongly on the model used.

To transform a measurement of the number counts from one filter to another, I first find a model which 
matches the counts in that filter reasonably well. When interpolating between two filters, such as between 
$I$ and $K$ to get the $H-$band number counts, the results are less model dependent when the same 
model fits the counts in both filters in which they have been measured. I run the model in the old and new 
filters, and determine the predicted median color as a function of magnitude in the first filter. I subtract 
the median color from the data magnitude, and determine the ratio of the measured counts to the 
predicted counts. I then multiply the model prediction for counts in the second filter by this ratio, and this 
is the prediction for the number counts in the second filter based upon the measured counts in the first 
filter. The error bars in $log(n)$ are the same. 

\begin{figure}

\plotone{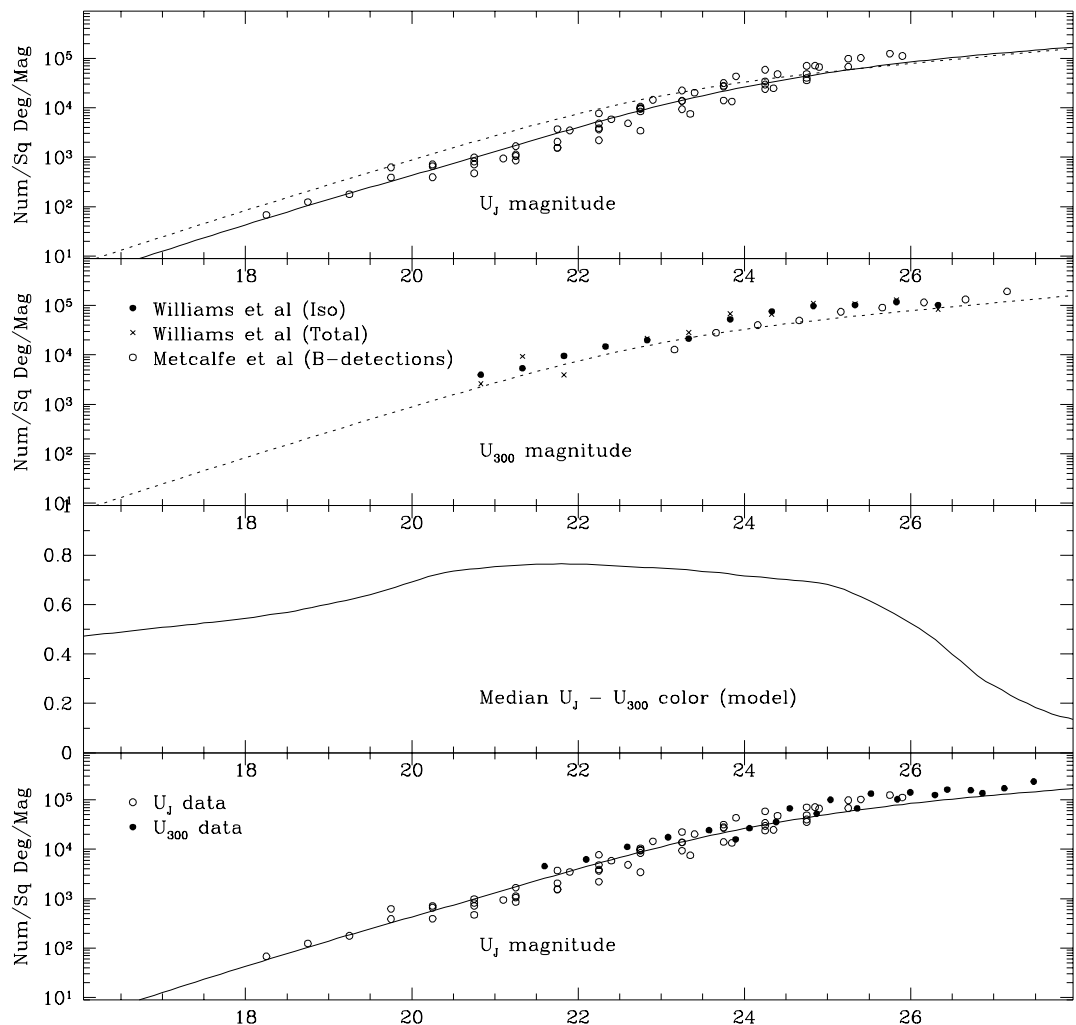}

\caption{The translation of number counts from the ground-based $U$ band to the $U_{300}$ filter. The 
top panel shows the ground-based $U-$band counts and a model which fits them (solid line), along with 
the model prediction for the $U_{300}$ counts. The next panel shows the HDF $U_{300}$ counts along 
with the model. The next panel shows the model's median $U_J - U_{300}$ color, which is then used to 
translate the $U_{300}$ counts into the $U_J$ filter, as explained in the text.}

\label{utrans}

\end{figure}

I will begin with a relatively simple example, converting the WFPC2 $U_{300}$ number counts in the 
HDF to the ground-based $U_{J}$ band, so that they can be plotted in the same figure. The top panel in 
Figure~\ref{utrans} is the ground-based $U_{J}$ counts plotted with a model for the $U_{J}$ counts 
(solid line), and the $U_{300}$ counts (dashed line.) To construct a model that fits the $U_{J}$ counts, I 
took a $q_0=0.02$ universe, included the effects of evolution and dust, and then set the mix of galaxy 
types to match the type-dependent results of the MDS (Glazebrook et al.\ \markcite{kgbmds}1995; Driver et al.\ \markcite{drivermds}1995) and the HDF 
(Abraham et al.\ \markcite{abraham}1996) in the $I-$band. This resulted in a late-type dominated galaxy mix with a higher overall 
normalization than the luminosity function measurements of Loveday et al.\ \markcite{loveday}(1992). Over the intermediate 
magnitudes, the model predicts different slopes for the two filters, as well as an offset of about $0.5mag$. 
The next panel in Figure~\ref{utrans} is the HDF $U_{300}$ number counts. I have plotted three 
versions of the counts, including the counts measured in isophotal and total magnitudes by 
Williams et al.\ \markcite{williams}(1996), and the counts measured by Metcalfe et al.\ \markcite{metcalfenat}(1996), who used aperture photometry in the 
$U_{300}$ image at the positions of objects identified on the $B_{450}$ image. The latter measurement 
goes deeper than the former, but is potentially subject to a selection effect against objects with extremely 
blue $U_{300} - B_{450}$ colors. The next panel in Figure~\ref{utrans} is the model prediction for the 
median $U_{J} - U_{300}$ as a function of $U_{J}$ magnitude. Finally, the $U_{300}$ counts have 
been transformed into $U_J$, and all the counts are plotted together.

To trace a single data point as an example, the Williams et al.\ \markcite{williams}(1996) measurement of the counts at 
$U_{300AB}=25.75$ is $log(n)=4.88 mag^{-1}~deg^{-2}$. Converting from the $AB$ magnitude 
system to the Vega magnitude system is an offset of $-1.42$ for the $U_{300}$ filter, which puts this 
point at $U_{300} = 24.33$. The median $U_{300} - U_{J}$ color at this point is $-0.71$, so the 
equivalent point is $U_J=25.04$. The model predicts that at $U_{300}=24.33$ the count will be $log(n) 
= 4.59 mag^{-1}~deg^{-2}$, which is a $0.29$ less than the measurement. The model predicts that the 
count at $U_J=25.04$ is $log(n)=4.71 mag^{-1}~deg^{-2}$. Thus the converted data point is 
$log(n)=5.00 mag^{-1}~deg^{-2}$ at $U_J=25.04$.

\begin{figure}

\plotone{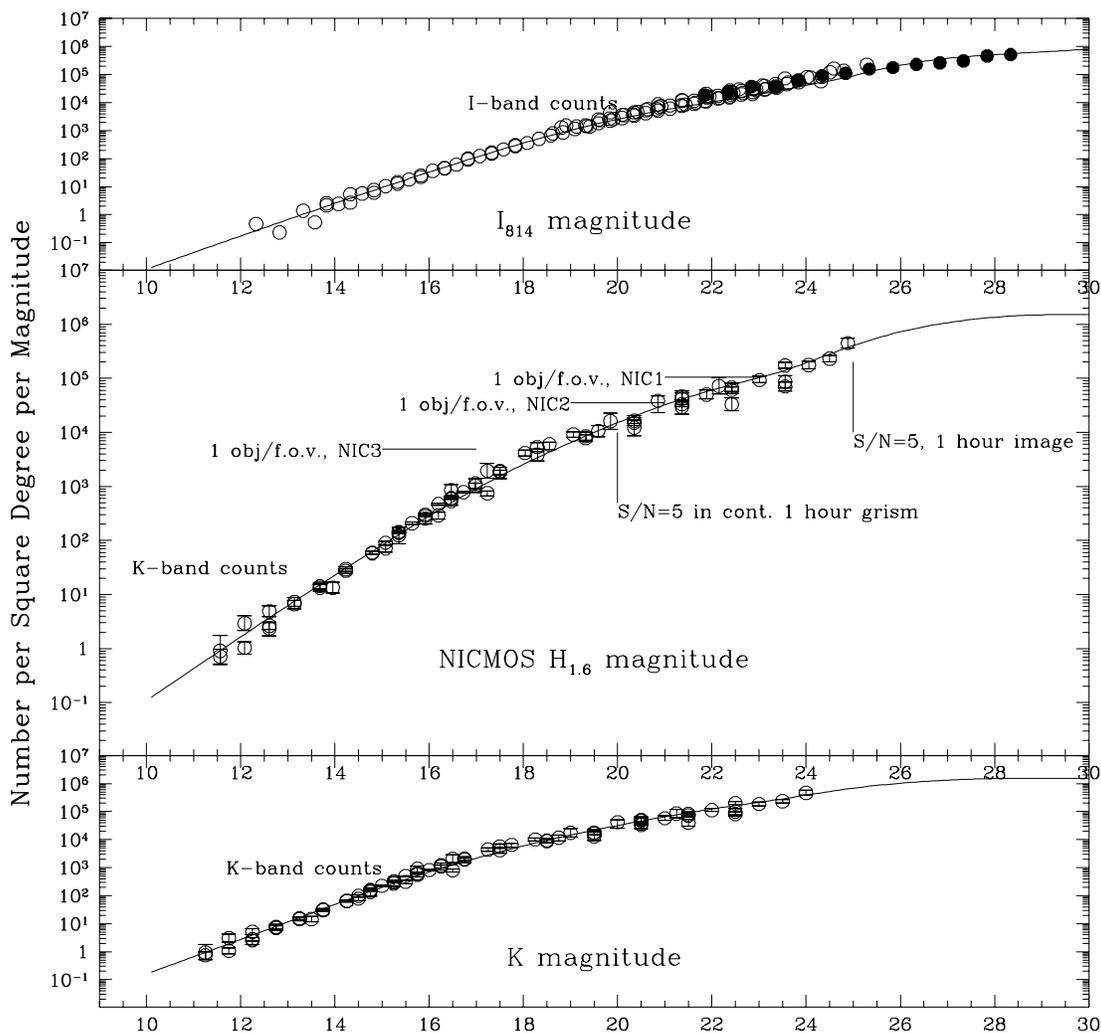}

\caption{The interpolation of number counts between the $I$ and $K$ bands, into the NICMOS 
$H_{1.6}$ filter. NICMOS is expected to reach S/N=5 for $H_{1.6}\approx 25$ for a point source in a 1 
hour exposure. This is well above the model prediction for 1 object per field of view with camera 2.}

\label{nich}

\end{figure}

I model the number counts in the NICMOS $H_{1.6}$ (F160W) filter in Figure~\ref{nich}. I have used 
the open universe model, including evolution and dust which gives a reasonable fit to the $I-$band 
number counts plotted in Figure~\ref{dustfig}. This model also fits the $K-$band number counts, and I 
have plotted the $I-$band counts in the top panel of the figure and the $K-$band counts in the bottom 
panel. The $K-$band counts have been transformed to the $H_{1.6}$ filter by the method discussed 
above, and the results are plotted in the middle panel. When background limited, NICMOS is expected to 
obtain a $5\sigma$ detection of point sources with $H_{1.6} = 25.0$ in a one hour exposure. 
Figure~\ref{nich} demonstrates that multi-orbit exposures with NICMOS will reach fainter magnitudes 
than current ground-based near-IR surveys.

\subsection{Extrapolation into the UV and Thermal IR}
 
The method described above can be used to extrapolate current observations into the UV and thermal IR, 
so long as appropriate choices of SEDs and evolutionary models are made. The GISSEL96 model extends 
into the UV and IR, although as discussed above it does not include the effects of dust internal to the 
galaxies. Dust absorbs the UV flux and re-radiates the energy in the mid-- and far--IR. While the simple 
dust model described in Section~\ref{dustsect} works reasonably well at modeling the effects on the UV 
flux from galaxies, the mid-- and far--IR is much more complicated. Galaxies have dust with a range of 
different temperatures, sizes and compositions, all of which affect the IR SED. In addition, starbursts due 
to merging events can dominate surveys conducted in the far--IR, and must be considered a separate 
population. The application of the techniques of this paper to a model of galaxy evolution in the far--IR 
will be considered in another paper. I consider here extrapolation of current observations into the UV, 
and into the mid--IR. While the mid--IR is also affected by some of the considerations discussed above, 
emission from faint galaxies at high redshift in the mid--IR comes from the rest-frame near--IR, a 
wavelength region well modeled by the GISSEL96 models. Thus the model is more appropriate for 
predicting the faint observations planned by SIRTF and NGST, than for predicting the brighter 
observations of ISO and WIRE.

\begin{figure}

\plotone{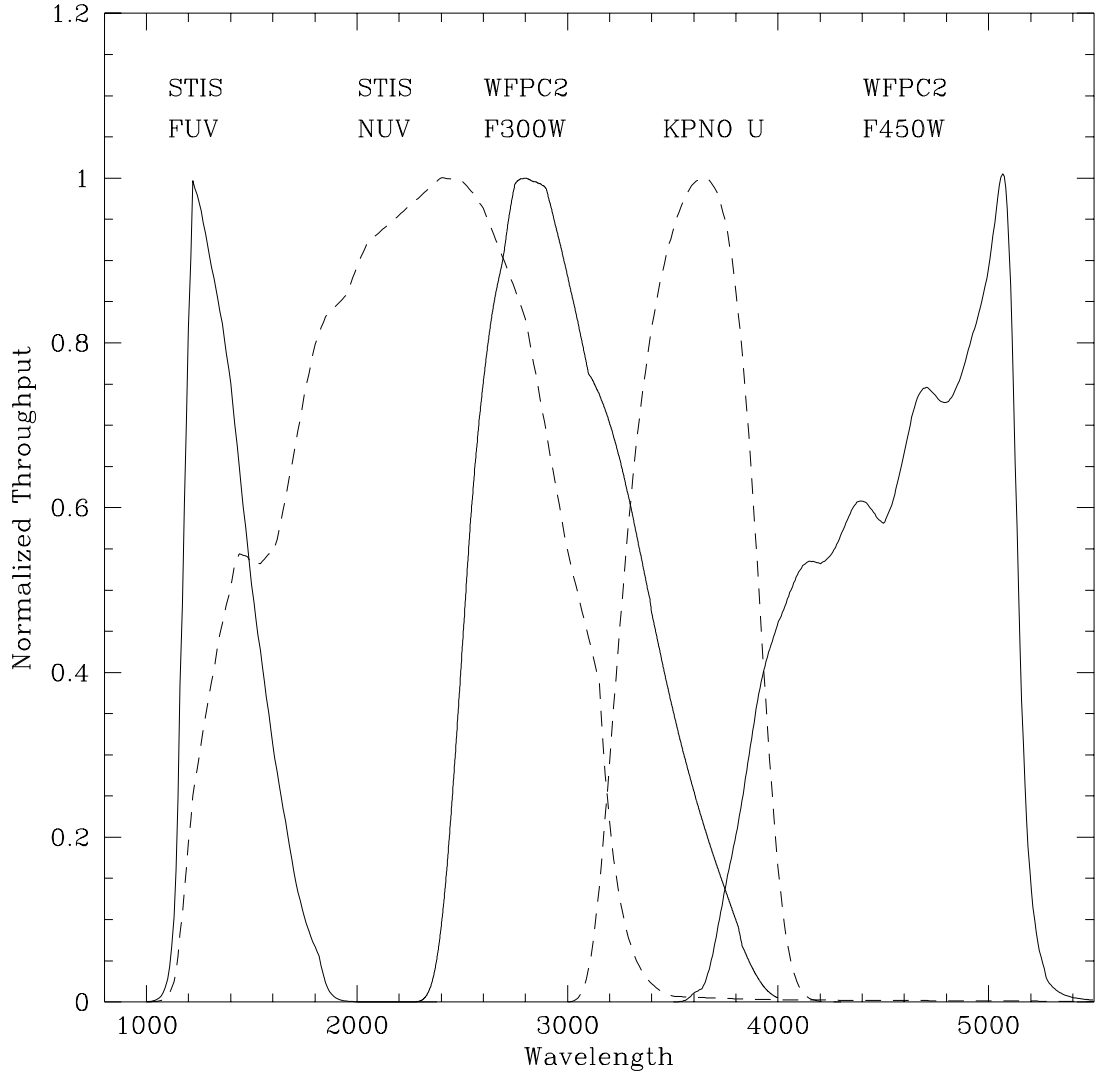}

\caption{Normalized filter tracings of various UV filter--detector combinations, including the STIS Far 
UV and Near UV MAMA detectors in clear mode, the WFPC2 $U_{300}$ and $B_{450}$ filters, and 
the ground-based $U$ filter.}

\label{ucomp}

\end{figure}

\begin{figure}

\plotone{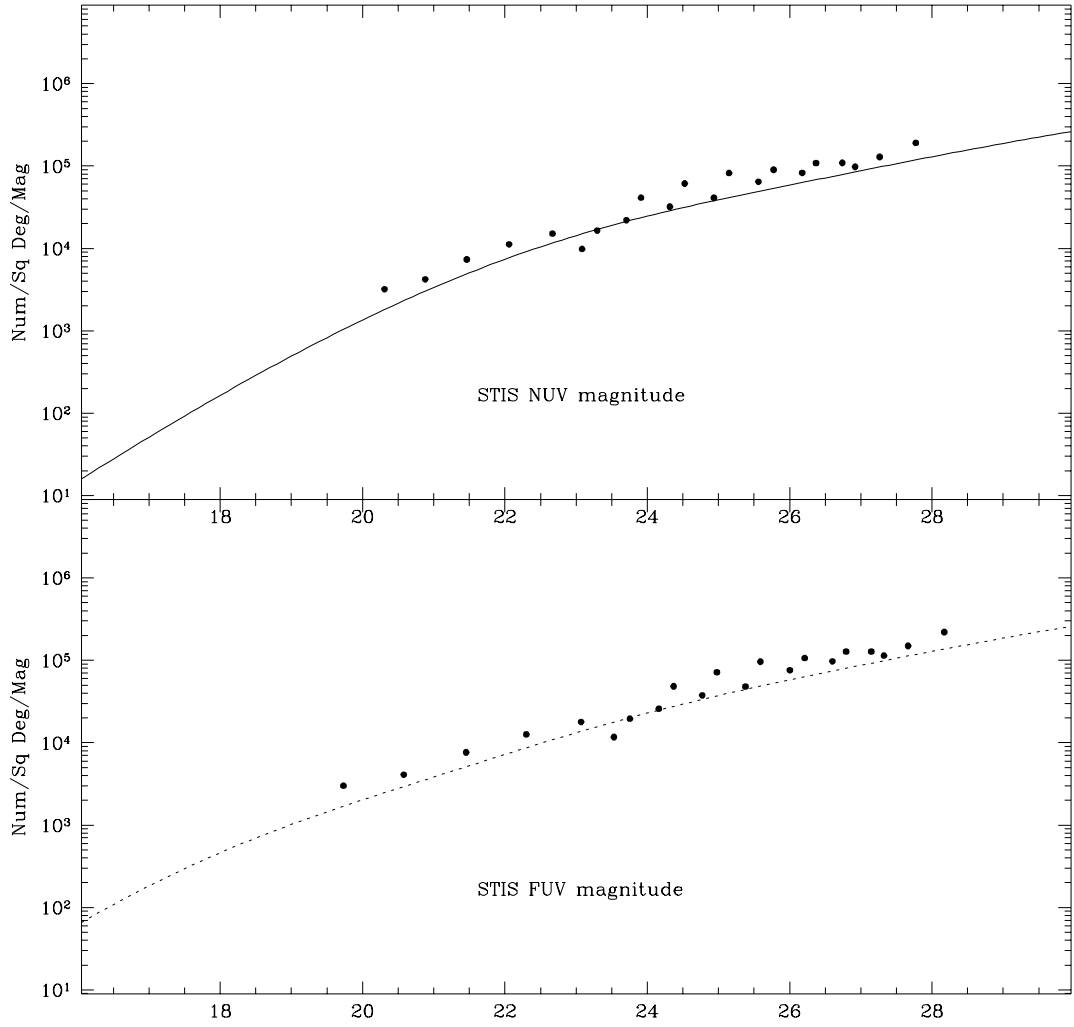}

\caption{A prediction of the number counts to be seen with STIS with the MAMA detectors in clear 
imaging mode.}

\label{stis}

\end{figure}

Figure~\ref{ucomp} shows normalized tracings of the STIS Near--UV and Far--UV MAMA detectors in 
their clear modes, compared to the WFPC2 $U_{300}$, ground-based $U_J$ and the WFPC2 $B_{450}$ 
filters. In Figure~\ref{stis} I have used ground-based $U-$band observations and the WFPC2 $U_{300}$ 
observations of the HDF to predict the number counts to be seen with STIS. Planned observations of an 
area within the Hubble Deep field using the STIS UV detectors are expected to reach fainter than $28 
mag$ in NUV, and $25 mag$ in FUV; the NUV exposure will reach fainter than the $U_{300}$ image 
and will strongly constrain galaxy evolution in the UV.

\begin{figure}

\plotone{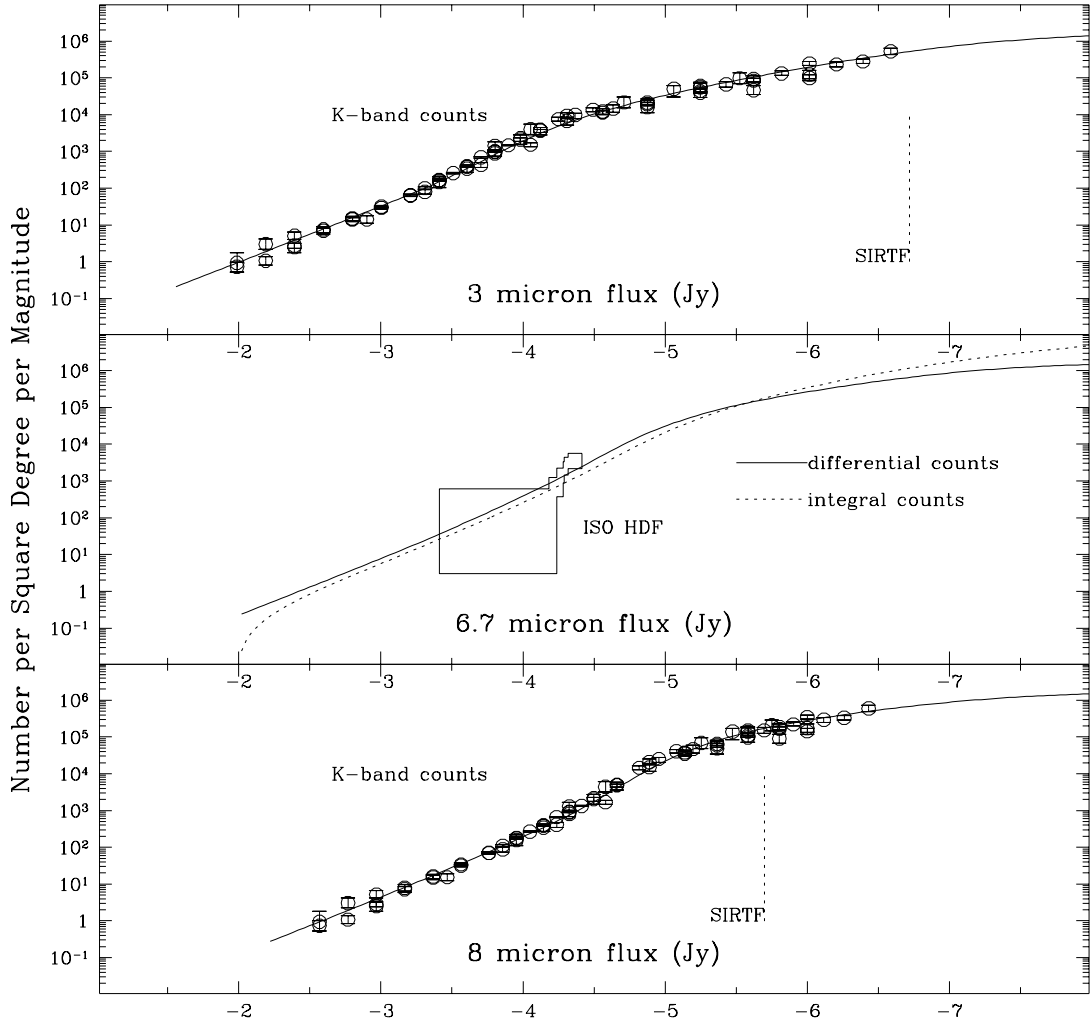}

\caption{A prediction of galaxy counts in the mid--IR. In the middle panel I plot the integral counts, and 
compare to the ISO observations on the Hubble Deep Field (Oliver et al.\ \protect\markcite{oliver}1997). These $6.7\mu m$ 
counts are only marginally inconsistent with the population of normal galaxies in my model, although the 
effects of dusty starburst galaxies, active galactic nuclei, and thermal re-radiation by dust are expected to 
become increasingly important in the mid--IR. Imaging with SIRTF is expected to become confusion 
limited at the sub--$\mu Jy$ level, while NGST is expected to achieve $nJy$ photometry.}

\label{sirtf}

\end{figure}

I plot an extrapolation of the $K-$band number counts into the mid--IR in Figure~\ref{sirtf}. Deep 
exposures at $3.5 \mu m$ and $8 \mu m$ are planned as part of the Legacy Science Program for the 
Space Infrared Telescope Facility (SIRTF) (Clemens, Greenhouse \& Thronson \markcite{clemens}1996). Observations of the HDF have been made at 
$6.7 \mu m$ with the Infrared Space Observatory (ISO) (Oliver et al.\ \markcite{oliver}1997), and the results are plotted in the 
middle panel of the figure. The planned observations by SIRTF will reach slightly deeper at $3.5 \mu m$ 
than ground-based $K-$band observations, although the planned observations at longer wavelengths will 
not. It should be noted, however, that an extrapolation in wavelength like I have made in this figure is 
model-dependent, and actual observations will constrain the models. In addition, of course, reaching 
fainter magnitudes or higher redshifts is not the only observational goal, and the mid--IR promises to play 
an important role in understanding the effects of dust and star-formation at low to intermediate redshifts. 
At the highest redshifts and faintest magnitude, the Next Generation Space Telescope, a 6 to 8 meter 
passively cooled instrument tentatively scheduled for launch around 2007, (Stockman \markcite{ngst}1997), will resolve and 
study the rest-frame UV, optical and near-IR at very high redshifts. 

\subsection{Color Distributions}

The color distribution of the galaxies in a survey holds more information about the state of evolution than 
the number counts. In the rest-frame, younger galaxies generally are bluer than older galaxies, but the 
K--correction can counter this effect. Comparing galaxies to evolutionary tracks without considering the 
number with each color will give an indication of redshift and evolutionary state for individual galaxies, 
but does not give a clear indication of the state of the entire population. Therefore, the color distribution 
as a function of magnitude must be modelled in order to interpret the data.

\begin{figure}

\plotone{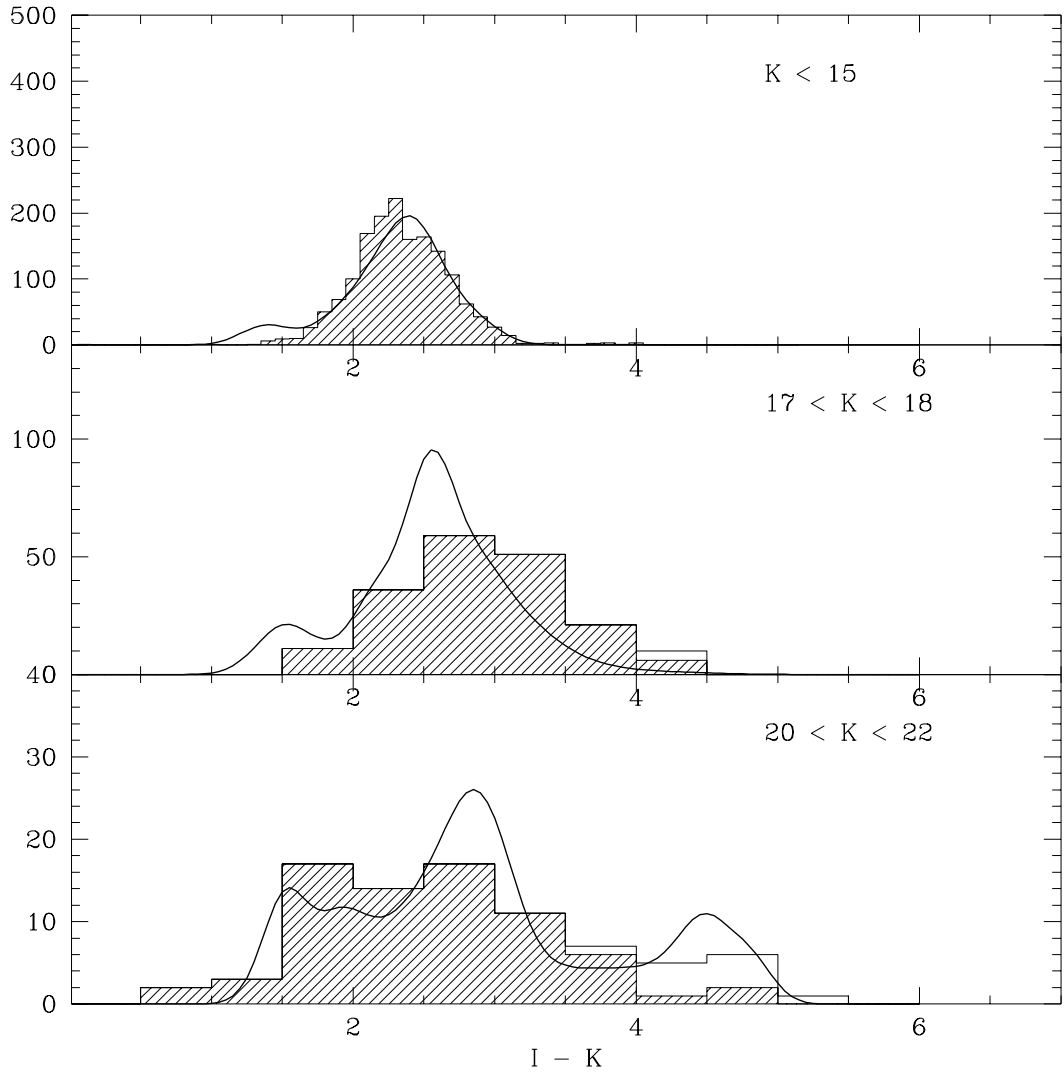}

\caption{A plot of the number of galaxies as a function of $I-K$ color. The model has been smoothed 
with a $0.2 mag$ Gaussian to mimic the effects of photometric noise. The dwarf galaxy model was used 
to increase the number of blue galaxies and dust was used to decrease the number of red galaxies.}

\label{colorplot}

\end{figure}

Figure~\ref{colorplot} is a plot of the number of galaxies as a function of $I-K$ color for three 
$K-$magnitude bins. The bright data are taken from Gardner et al.\ (in preparation) and the fainter data 
are taken from Gardner \markcite{gardner95}(1995). The models plotted in Gardner \markcite{gardner95}(1995) were based upon the original 
Yoshii \& Takahara \markcite{yt88}(1988) models which do not include the effects of dust. The primary effect of dust in this figure is to 
block the bright ultraviolet flux from rapidly starforming galaxies at high redshift which come to 
dominate the population at faint levels. Galaxies appear red in the models because of their high redshift. 
Model predictions of the colors of galaxies are often not smooth functions because of the discrete nature 
of the models, so in this figure I have convolved the output with a $0.2 mag$ Gaussian filter to mimic the 
effects of photometric noise. While it is always possible to run the model with a higher resolution in 
redshift or magnitude, models usually use discrete types. In the model described here, I have removed 
this restriction, and allowed interpolation between the different integral types. To fit the data I plot an 
open universe model with evolution, dust and dwarf galaxies. This figure is an indication of what this 
model can do, but the interpretation of color distributions is more subject to uncertainties in the free 
parameters than are number counts.

\subsection{Luminosity Function Conversion}

The three parameters of the Schechter luminosity function (LF) are highly correlated, and any 
comparison of luminosity functions measured in different surveys must take this correlation into account. 
The maximum likelihood method for determining a parametric LF from a field galaxy redshift survey 
without binning the data was originated by Sandage, Tammann \& Yahil \markcite{sty}(1979). This method has the advantage over previous 
methods, (see Felten  \markcite{felten}1977 for a review), that it is relatively insensitive to the effects of clustering, using 
the assumption that the LF is independent of environment. The maximum likelihood method, however, is 
subject to a normalization constant that must be determined in another way, so it returns $M^*$ and 
$\alpha$, and the error ellipse for those two parameters, while $\phi^*$ must be determined 
independently. It is possible to determine $\phi^*$ by normalizing a model of the galaxy counts based 
upon a measured $M^*$ and $\alpha$ to observed bright galaxy counts (Mobasher, Sharples \& Ellis \markcite{mobklf}1993; 
Gardner et al.\ \markcite{gardnerklf}1997). Unless model counts are run for every point within the error ellipse in $M^*$ and 
$\alpha$, however, this method does not return the correlation between the errors in all three parameters.

To compare LFs determined in different filters, it is best to compare $M^*$ and $\alpha$ with a 
translation between the filters, and then to compare $\phi^*$ by comparing the bright galaxy counts 
directly with the galaxy count model. The galaxy count model presented in this paper converts LFs 
between filters using a method developed by Yoshii \& Takahara \markcite{yt88}(1988). The type-independent or type-dependent values 
of $M^*$ are converted by adding the model rest-frame color, and $\alpha$ is assumed to be the same for 
each filter. This will sometimes have the effect of converting a type-independent LF into a 
type-dependent LF, when the rest-frame colors are type-dependent. To compare the error ellipses directly 
would require doing this procedure for every point within the error ellipse. However, since the sum of 
several Schechter LFs (i.e. one for each galaxy type used in the model) is not likely to equal a LF which 
is easily parameterizable as a Schechter LF, this is not a straight-forward task. It is much easier to 
compare LFs graphically.

\begin{figure}

\plotone{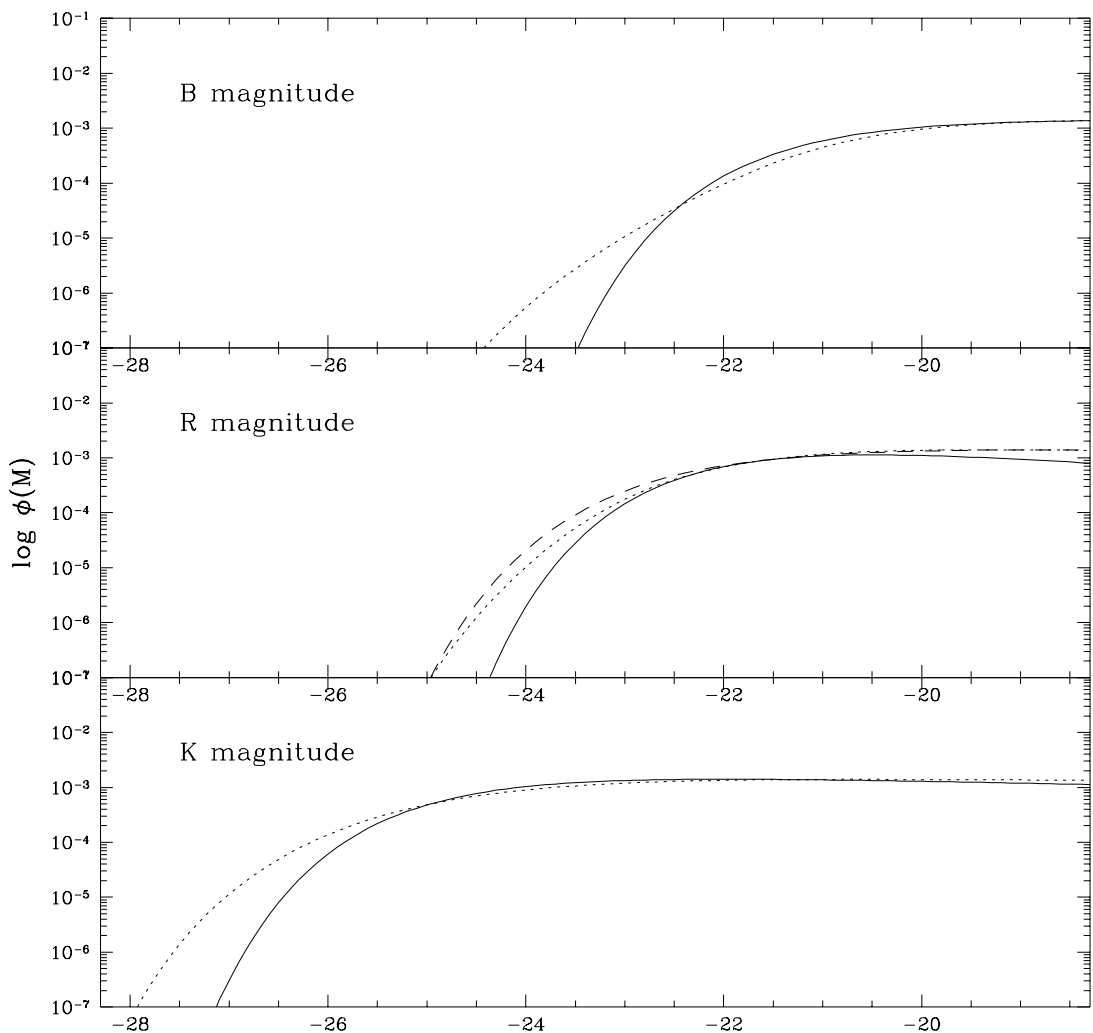}

\caption{Comparison of luminosity functions determined in the $b_J-$band, the $R-$band, and the 
$K-$band. The top panel contains the $b_J-$band LF of Loveday et al.\ \protect\markcite{loveday}(1992) with the $K-$band LF of 
Gardner et al.\ \protect\markcite{gardnerklf}(1997) converted to the $b_J$ band using the methods described in the text. The middle 
panel contains the $R-$band LF from Lin et al.\ \protect\markcite{lin}(1996) with the $b_J-$band and $K-$band LFs converted 
to the $R$ band. The bottom panel contains the $K-$band LF of Gardner et al.\ \protect\markcite{gardnerklf}(1997), along with the 
$b_J-$band LF converted to the $K$ band.
}

\label{lfcomp}

\end{figure}

The top panel in Figure~\ref{lfcomp} is a comparison of the $b_J-$band luminosity function measured 
by Loveday et al.\ \markcite{loveday}(1992), with the $K-$band LF measured by Gardner et al.\ \markcite{gardnerklf}(1997), converted to the $b_J-$band 
using this technique. The excess of bright galaxies in this figure is due to the assumption that $M^*_K$ 
is independent of type, causing the bluest galaxies to appear too bright in $b_J$. This assumption has 
little effect on model predictions of the near-infrared properties of galaxies due to the low number of blue 
galaxies, but is magnified by the conversion of the LF to the blue filter. It is an indication of the 
uncertainties introduced into this type of model by the extrapolation over a factor of five in wavelength. 
The middle panel contains the measured $R-$band LF of Lin et al.\ \markcite{lin}(1996), and the converted $b_{J}-$band and 
$K-$band LFs. The bottom panel contains the $K-$band LF and the $b_J-$band LF converted into $K$. 
The assumption that $M^*_{b_J}$ is independent of type is more accurate than in $K$.

\section{Summary}

Galaxy counts are one of the classical cosmological tests, but their interpretation remain difficult. Model 
predictions of the counts, and the color and redshift distributions of galaxies are subject to uncertainties 
in the spectral energy distributions and evolution of galaxies, and in free parameters specifying the 
luminosity functions, cosmological geometry, the number and distribution of galaxy types, and the effects 
of dust and merging. However, there is a growing body of observational data that constrains these 
models. In this paper, I have presented a model in which the free parameters can be easily adjusted, and  
which can be extended to include new parameters. The greatest power of this model is in converting 
observational data taken in one filter into another filter to compare with other data, and extrapolating to 
longer or shorter wavelengths.

The software, parameter files and data presented in this paper are available at 
http://hires.gsfc.nasa.gov/$\sim$gardner/ncmod.

\bigskip

I wish to acknowledge funding by the Space Telescope Imaging Spectrograph Investigation Definition 
Team through the National Optical Astronomical Observatories, and by the Goddard Space Flight Center.

\clearpage

\clearpage

\clearpage

\clearpage

\clearpage

\clearpage

\clearpage

\clearpage

\clearpage

\clearpage

\clearpage

\clearpage

\clearpage

\end{document}